# Repeated Bifurcation of Relativistic Magnetic Pulse and Cosmic Gamma-Ray Bursts

Edison Liang

*Rice University, Houston, TX 77005-1892*

and Kazumi Nishimura

*Los Alamos National Laboratory, Los Alamos, NM 87545*

The diverse and complex light curves of gamma-ray bursts (GRBs) remain an outstanding astrophysical mystery. Here we report the results of 2-1/2-dimensional particle-in-cell (PIC) simulations of the relativistic expansion of magnetized electron-positron plasmas. When the simulation is carried to >150 light-crossing time of the initial plasma, the plasma pulse reproduces many of the GRB features. Remarkably, the plasma pulse bifurcates repeatedly, leading to a complex, multi-peak structure at late times, resembling GRB profiles.

PACS numbers: 52.65 Rr 52.30.-q 52.65.-y 95.30Qd

Gamma-ray bursts (GRBs) [1-2] exhibit a number of distinctive features that have defied explanation. Among them are the diverse and complex light curves [1], unique spectra and spectral evolution [2]. In addition the gamma radiation and particle energization mechanisms have not been identified. The recent discovery of a strongly polarized GRB [3] supports the presence of strong, ordered magnetic field at the source. Here we report results from PIC simulations [4-5] of the expansion of magnetized electron-positron plasmas potentially relevant to the temporal and spectral properties of GRBs. The major new results from these simulations are: (a) repeated bifurcation of the plasma pulse at late-times, (b) development of a power-law with low-energy cutoff in the particle momentum distribution, (c) a square-root scaling law for the average Lorentz factor.

When a hot, $\beta$ (= thermal pressure/magnetic pressure) $\leq 1$, collisionless plasma with a transverse magnetic field expands into a vacuum or low-density surrounding, the self-induced transverse currents [6] reshape the electromagnetic (EM) pulse in such a way that it traps and accelerates the surface particles via the pondermotive force [7]. For a $\beta \sim 1$ e+e- plasma, this mechanism, called the diamagnetic relativistic pulse accelerator (DRPA) [8], converts most of the initial magnetic energy into the ultra-relativistic directed energy of a fraction of the surface particles. However previous simulations [8] stop at t<10 $L_o/c$ ($L_o$= initial plasma thickness, c=light speed), unable to reveal the late-time behavior. The present simulations are carried to t>150 $L_o/c$. Fig.1 highlights the global evolution of a diamagnetic relativistic pulse (DRP) expanding into a vacuum in both slab and cylindrical geometries. At late-times, the EM peak width $\Delta_B$ approaches $\gamma_m c/\Omega_e(t)$, where $\Omega_e(t)$ is the instantaneous gyrofrequency [9] at the pulse peak (throughout this paper all variables refer to the laboratory frame) and $\gamma_m$ is the average Lorentz factor of the surface particles. In the following we focus on the slab results since the cylindrical simulation cannot yet achieve the same resolution as the slab case.

As the pulse advances, it bifurcates (Fig.2) due to the formation of new traps in the EM field. This bifurcation process begins at t~10.$L_o/c$ and repeats indefinitely, leading to a complex multi-peak structure at late times. We can visualize the detected light curve emitted by the particles of Fig.2 as follows. Particles moving at high

Lorentz factor $\gamma$ with the same x(t) emit photons that arrive at the detector at almost the same time due to the time compression factor [9] $1/(2\gamma^2)$. Yet photons emitted by particles at x(t)-$\Delta$x are delayed by $\Delta x/c$, independent of $\gamma$. Let $t_r$ be the moment when radiation loss dominates (see below). If the particles radiate away their energy in $\Delta t_r \ll \gamma^2 L(t_r)/c$ (L(t)=density pulse width at t), then the detected luminosity light curve traces the energy density profile $n(x,t_r)<cp_x(x,t_r)>$ (n=particle density, $<p_x(x)>$=average momentum at x). Since the average photon energy emitted by particles of momentum $<p_x>$ is $\propto <p_x>$ (see below), the photon number light curve mimics the snapshot of $n(x,t_r)$ to first order.

DRPA thus provides a plausible explanation for the diversity of GRB light curves. If a DRP radiates all its energy before bifurcation (Fig.2a), it produces a smooth single-peaked GRB (i.e. a Fast Rise Exponential Decay pulse or FRED [1]). If it radiates only after repeated bifurcation (Fig.2d-e), it produces a GRB with complex, multiple peaks. The radiation time $t_r$ and $\Delta t_r$ depend on the initial conditions and ambient photon and plasma density (see below). In reality, the mapping between $n(x,t_r)$ and the detected light curve also depends on the detector threshold (very soft photons are undetected), view angle, 3-D effects and potentially different $t_r$ at different parts of the pulse. GRBs with multiple FRED-like [1] smooth pulses and GRBs with long periods of non-emission [1] may involve multiple DRP's or ion-loaded plasmas [8, 10]. Previous results on pure e-ion plasma expansion [8,10] show that ions get most of the EM pulse energy due to charge separation. But we speculate that this would not happen when a small amount of ions is mixed into a e+e- plasma, because the charge separation electric field does not act on the neutral e+e-component. We expect the EM pulse to preferentially trap and accelerate the more mobile e+e- plasma, leaving the e-ion plasma behind.

Other unique properties of GRBs are also reproduced in Fig.2. (a) The "hard-to-soft" spectral evolution of most GRBs, especially FRED's [2], is evident from the phase plots of Fig.2. (b) The well-known rise-decay asymmetry of FRED-like pulses [11-12] is evident in Fig.2a. Our simulated pre-bifurcated pulses have rise-decay asymmetry ratios ~ 0.5 at FWHM, similar to those of FREDs [11-12]. (c) Narrow peaks in complex GRBs tend to be more symmetric than FREDs [12]. This is also evident in Figs.2b-2e. Simulations with additional nonaxial B-components and with ambient plasmas produce more diverse bifurcation patterns that further account for the variety of GRB profiles. Fig.3 shows pulses from runs with nonzero $B_z$ and nonzero ambient plasma density. These results confirm that, (a) the DRPA mechanism is robust and not inhibited by nonaxial components of B that couples axial and radial motions, or by the early interaction with cold ambient plasma, (b) the maximum Lorentz factors achieved in these cases are comparable to the benchmark of Fig.2a, (c) there is no 2-D plasma instability at the DRP interface with ambient cold plasma. Such instabilities (e.g. 2-stream) may be suppressed by the strong transverse EM field of the pulse. Figs.3bc also show that when a DRP interacts with dense ambient plasma, it generates "soft precursors" of swept-up ambient plasma. This may be relevant to GRBs with soft x-ray precursors [1].

As the DRP advances, a peak develops in the momentum distribution (Fig.4) that may explain the origin of the ubiquitous GRB spectral break [2]. Particles whose x-momenta lie below the peak $p_{xmax}$, which corresponds to the group Lorentz factor of the EM pulse, gradually fall behind the pulse and lose acceleration, creating a deficit of low-energy particles in the pulse (Fig.2 phase plots). At $\Omega_e.t > 5000$ a power law develops above $p_{xmax}$. In this example the power-law slope of -3.5 translates into a photon index [13] of 2.25, similar to the index of GRBs [2]. Studies suggest that the the power-law high-energy cutoff $\gamma_{lim}$ is limited by $L(t)\Omega_e(t)/c$ as the pulse width $L(t)$ limits the maximum relativistic gyroradius $c\gamma_{lim}/\Omega_e(t)$.

The most important result emerging from these long-duration simulations is the growth of the peak Lorentz factor $\gamma_m(t)$ ($=p_{xmax}/m_e c$) with t according to

$$\gamma_m(t) = (2.f.\Omega_e(t).t + C_o)^{1/2} \qquad t > L_o/c \qquad (1)$$

where $C_o$ and f are constants dependent on initial conditions (Fig.5). Eq.(1) can be derived using the phase-averaged Lorentz equation for particles comoving with the EM pulse [9]. Assuming that Eq.(1) applies to cosmic dimensions, this simple square-root scaling allows us to check if DRPA is consistent with GRB parameters. Typical long GRB duration is $L(t_r)/c \sim 30$ seconds [1]. As an example we take $t_r = 300$ sec, since most GRBs have already bifurcated when they radiate ($t_r \geq 10.L_o/c$). The characteristic frequency of DRPA radiation can be estimated dimensionally using [9]: $\omega_{cr} \sim \gamma^2 c/\Delta_B \sim \gamma_m \Omega_e(t_r)$ (note the linear dependence on $\gamma_m$). Eq.(1) plus $h\omega_{cr}/2\pi \sim 500$ keV ([2] "deredshifted") give $\gamma_m \sim$ few.$10^7$ and $B \sim 10^6$ G at $t_r$. Internal shock models [14] require bulk Lorentz factor $\Gamma \sim 10^2$ plus internal (isotropic) Lorentz factor $\gamma_{int} > 10^4$ for synchrotron emission, giving a composite Lorentz factor up to $>10^6$. But the two cannot be directly compared since the DRPA $\gamma_m$ is unidirectional. However, spherical divergence, interaction with dense ISM and Compton loss [9] against ambient photons may reduce $\gamma_m$ to values below Eq.(1). Such environmental effects may result in x-ray-dominated GRBs [1-2].

Preliminary results hint that the minimum width of the bifurcated peaks scales as the geometric mean of $c\gamma_m(t)/\Omega_e(t)$ and pulse length $L(t) \sim c\gamma_{lim}/\Omega_e(t)$, but this remains to be confirmed. Using the above numbers this suggests a minimum subpulse duration of ~10 ms. Also $B \sim 10^6$ G gives a magnetic energy $E_B(t_r) \sim 10^{50}$ ergs, assuming a $4\pi$ shell of $L(t_r) \sim 10^{12}$ cm and radius $R \sim ct_r \sim 10^{13}$ cm. This gives a total initial energy [8] $E_{tot} \sim 10 E_B \sim 10^{51}$ ergs, consistent with typical GRBs. Magnetic flux is not conserved in a DRP expansion due to current generation. But the sum of the magnetic and particle energy is conserved, before radiation loss. Hence if the DRP originates from a region $<10^7$ cm and initial $\beta < 1$, we need an initial $B >$ few.$10^{15}$ G, hinting at a magnetar connection.

The EM pulses with large transverse current emergent from our initially confined magnetic field may be symptomatic of more generic magnetic-dominated (Poynting flux) outflows [15] capable of DRPA-like action. Poynting flux models are favored if the recent RHESSI polarization result [3] is confirmed for other GRBs, since internal shock models [14] have difficulty generating large-scale ordered magnetic fields. We speculate that any magnetic-dominated mildly relativistic plasma that is suddenly "deconfined", may generate EM pulses similar to the DRP. A relevant example is a rising flux rope generated by the strongly magnetized accretion disk of a newly formed blackhole in the center of a hypernova [16]. When the flux rope emerges from the stellar surface, its expansion into the surrounding low density environment may mimic the DRP expansion. As the next step in our simulations we will study the effect of turning off the surface current of a confined magnetic field over different time scales, instead of instantaneously, in 2-and-3-D geometries.

E.L. was supported by NASA grant NAG5-7980 and LLNL contract B510243. K.N. was supported by the LANL LDRD program under the auspices of the US DOE.

---

Figure Captions

Figure 1  2-1/2-D PIC simulations of slab and cylindrical magnetized relativistic e+e- plasma expansion, with initial plasma temperature kT=5 MeV, $\Omega_e/\omega_{pe}$=10 ($\Omega_e$=$eB_o/cm_e$, $\omega_{pe}$=$(ne^2/\pi m_e)^{1/2}$), initial slab width $L_o$=120$c/\Omega_e$ and uniform internal **B**=(0, $B_o$, 0).  We show the x≥0 snapshots of particle distribution (a-d), axial magnetic field (color scale runs from $B_y$=+0.2$B_o$ (red) to -0.1$B_o$(blue)) and current density (white arrows) for the cylindrical case (e), and phase plot for the slab case (f). $\Omega_e t$=800 for all left panels and $\Omega_e t$=$10^4$ for all right panels.  The green dot in the phase plot denotes the initial phase volume.  Results for x<0 are identical.

Figure 2  Particle density profiles (blue curves, right scales) and phase plots (red dots, left scales) for the slab run of Fig.1 at (a)$\Omega_e t$=1000, (b)5000, (c)10000 and (d)18000, with current densities in small insets.  Panel (e) is the $\Omega_e t$=30000 snapshot of another run with $L_o$=600$c/\Omega_e$, showing more peaks.  These results should be compared to the GRB light curves of [1].  The "hard-to-soft" trend of the momentum distribution is clearest in panel (a), where the momentum peaks before the density, in agreement with BATSE data for FREDs [1].

Figure 3  Particle density profiles (blue curves, right scales) and phase plots (red dots, left scales) at $\Omega_e t$=1000 for runs with (a) $B_z$ increasing linearly from $B_z$=0 at x=0 to $B_z$= $B_o$ (-$B_o$) at x=$L_o$/2 (-$L_o$/2), (b) cold ambient e+e- density = 5% of slab density, (c) cold ambient density = 20% of slab density.  Other initial conditions are identical to Fig.2a.

Figure 4  Evolution of the x-momentum distribution for all surface particles in the slab pulse of Fig.1, showing the development of the peak Lorentz factor $\gamma_m$(=$p_{xmax}/m_e c$). and a power law with slope ~ –3.5.

Figure 5 The peak Lorentz factor $\gamma_m$ versus time for the slab pulse, compared with Eq.(1). The best-fit curve (dotted) gives f=1.33 and $C_o$=27.9. $\Omega_e(t).t$ =3800 is equivalent to $\Omega_e.t$=18000 due to B decay.

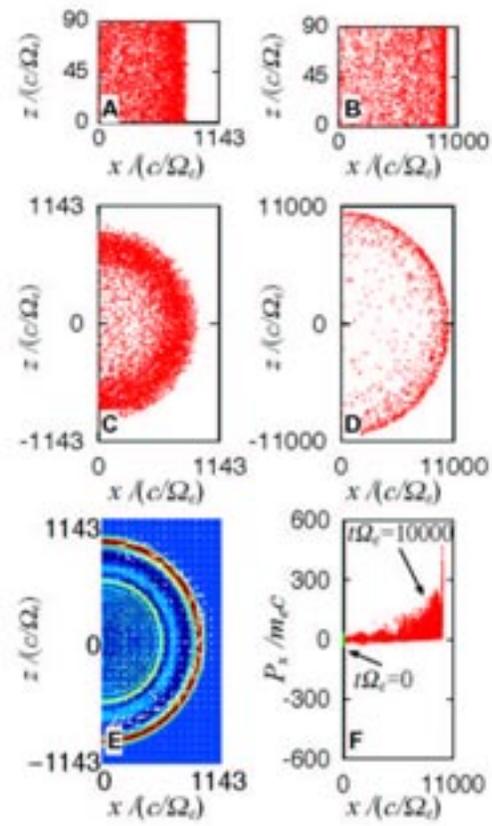

Fig.1

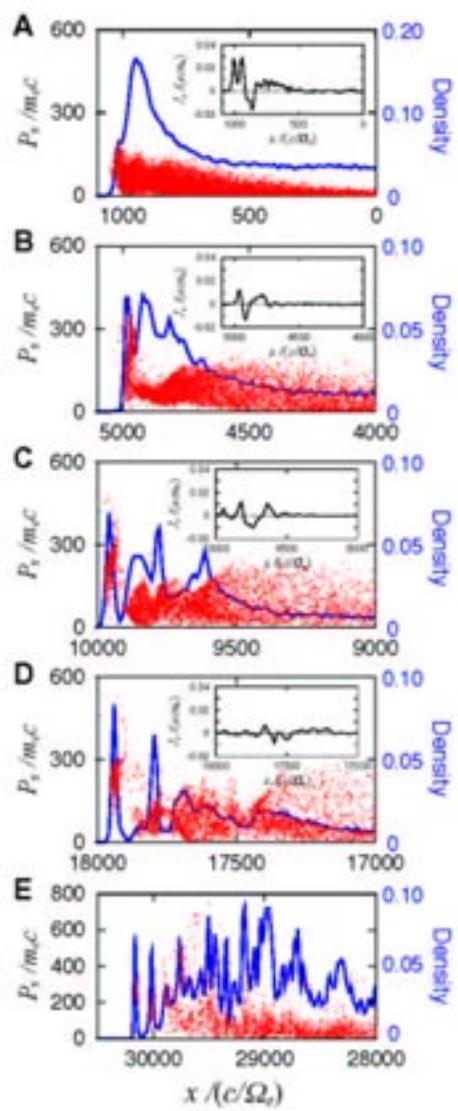

Fig.2

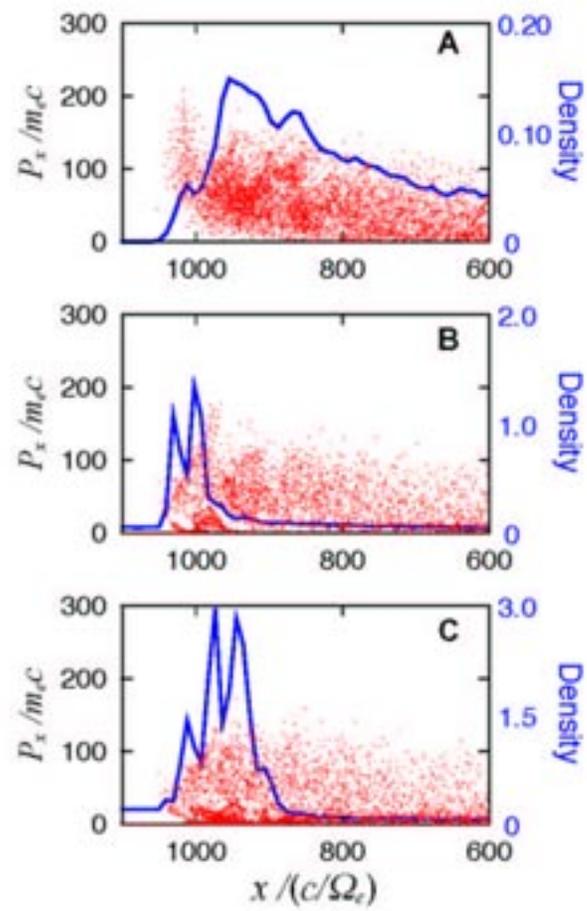

Fig.3

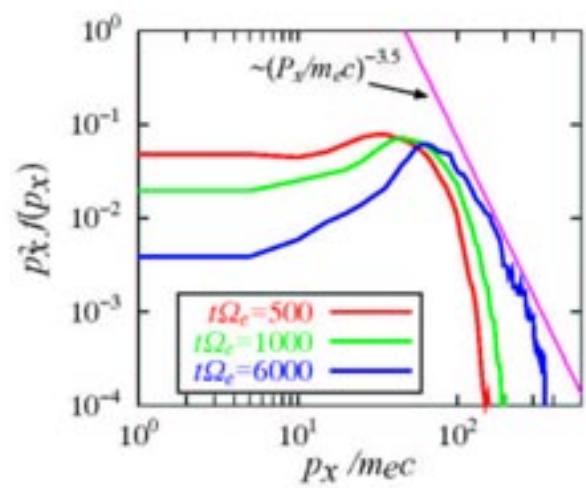

Fig.4

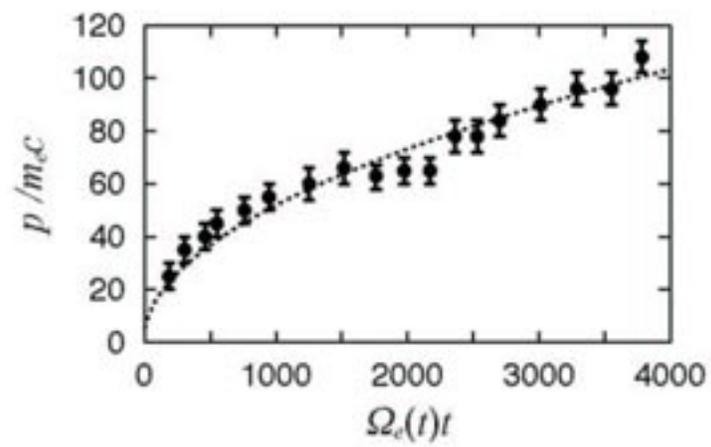

Fig.5